# Rare semileptonic decay of $\chi_{c1}(1p)$ meson

# In QCD

N. Ghahramany, A. R. Houshyar

Physics Department, College of Sciences, Shiraz University

and Biruni observatory, Shiraz 71454, Iran

## Abstract

The rare semileptonic $\chi_{c1}(1p) \to D_s^+ e \bar{\nu}$ decay is analyzed, by using the three-point QCD sum rules. Taking into account the two-gluon condensate contributions, the transition form factors related to this decay are calculated and are used to determine the total decay width and branching fraction. Our findings may be approved by future experiments.

# 1 : INTRODUCTION

Quarkonia are bound states of $Q\bar{Q}$, where $Q$ is a heavy quark, either a charm quark, $c\bar{c}$ (charmonium) or a beauty quark, $b\bar{b}$ (bottomonium). Toponium does not exist, since the top quark decays through the electroweak interaction before a bound state can take form. In case of the lighter quarks (u, d, s ), the physical states seen in experiments are quantum mechanical mixtures of the light quark states. The analysis of heavy quark and anti-quark systems are proper candidate for applying QCD. Production and decays of quarkonium have long been used to investigate the nature of QCD. Due to heavy, but not very heavy quark mass, one can get knowledge of both perturbative and non-perturbative QCD through the analysis of the natures of production and decays of quarkonium. Properties of these systems have been theoretically calculated, mainly using potential model, where the potential, $V = -\frac{4}{3}\frac{\alpha_s}{r} + kr$, describes the static potential of the quarkonia, or its extension like the coulomb gauge model [1-5]. The first term in the above potential is related to one gluon exchange and the second term is called the confinement potential.

A recent study on the calculation of ground-state decay constant by QCD sum rules and potential models show that they both follow the same pattern. In addition, it has been revealed that using QCD sum rules, we can get a more precise calculation of the bound-state characteristics as compared to the potential models [6]. The QCD sum rules are a reliable method for spectroscopy and obtaining the properties of the hadrons [7-10]. Furthermore, this method has been used for the calculation of masses and decay constants of the mesons [11-18]. Semileptonic decay of heavy mesons has been the aim of many recent studies. Semileptonic decay of the scalar, pseudoscalar, vector and axial vector mesons using three-point QCD sum rules, were the subject of these papers [19-29].

The present research was undertaken to study the semileptonic decay of axial vector p-wave charmonium $\chi_{c1}(1p)$ with the quantum numbers $J^{PC} = 1^{++}$ into pseudoscalar $D_s^+$ meson. The objective of this work is to evaluate the decay width of $\chi_{c1}(1p) \rightarrow D_s^+ e\bar{\nu}$ by considering two gluon condensates as the first non-perturbative contribution to the correlation function. Heavy quark condensates are suppressed here by the inverse powers of the heavy quark mass.

This paper includes the following sections; in section two, the calculation of sum rules is presented for the related form factors in which the two gluon condensates contributions to the correlation function is considered. Section three consists of the numerical analysis of the form factors and the estimation of the branching fraction. In the last section the conclusions are discussed.

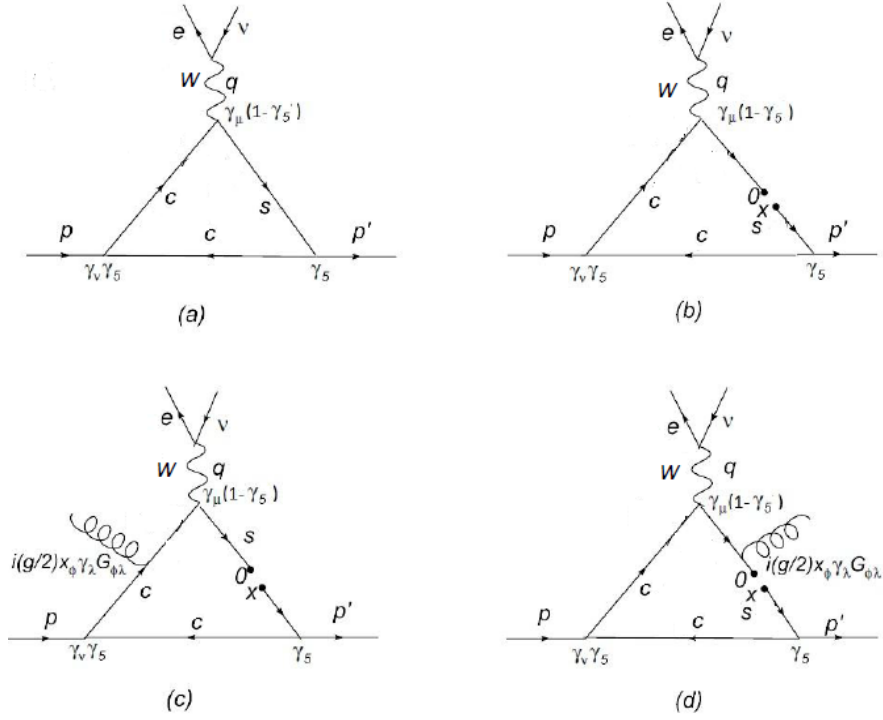

Figure 1: (a) Bare loop. (b) Light quark condensates. (c, d) Light quark condensates with one gluon emission for $\chi_{c1}(1p) \to D_s^+ e \bar{\nu}$ decay.

## 2 : Theoretical analysis of form factors for $\chi_{c1}(1p) \to D_s^+ e \bar{\nu}$ in the context of QCD Sum rules

The decay of $\chi_{c1}(1p) \to D_s^+ e \bar{\nu}$ is described in the standard model, in terms of quark degrees of freedom by the process $c \to s e \bar{\nu}$ at tree-level (Fig. 1). The effective Hamiltonian for this transition can be written as:

$$H_{eff} = \frac{G_F}{\sqrt{2}} V_{cs} \bar{\nu} \gamma^\mu (1 - \gamma_5) e \bar{s} \gamma_\mu (1 - \gamma_5) c, \qquad (1)$$

where $G_F$ stands for Fermi constant and $V_{cs}$ is the element of the Cabibbo-Kobayashi-Maskawa (CKM) matrix. The transition amplitude of the $\chi_{c1}(1p) \to D_s^+ e \bar{\nu}$ is obtained by sandwiching Eq. (1) between the initial and final meson states,

$$M = \frac{G_F}{\sqrt{2}} V_{cs} \bar{\psi}_{\bar{\nu}} \gamma^\mu (1 - \gamma_5) \psi_e \langle D_s^+(p')|\bar{s}\gamma_\mu(1-\gamma_5)c|\chi_{c1}(1p)(p,\varepsilon)\rangle \qquad (2)$$

To continue, we need to calculate the matrix element $\langle D_s^+(p')|\bar{s}\gamma_\mu(1-\gamma_5)b|\chi_{c1}(1p)(p,\varepsilon)\rangle$. We parameterize the matrix element in terms of the form factors:

$$\langle D_s^+(p')|\bar{s}\gamma_\mu(1-\gamma_5)c|\chi_{c1}(1p)(p,\varepsilon)\rangle = -\epsilon_{\mu\nu\alpha\beta}\,\varepsilon^{*\nu}p^\alpha p'^\beta \frac{2V(q^2)}{m_{D_s^+}+m_{\chi_{c1}(1p)}}$$
$$+i[\varepsilon^*_\mu(m_{D_s^+}+m_{\chi_{c1}(1p)})A_1(q^2) -$$
$$(\varepsilon^* q)P_\mu \frac{A_2(q^2)}{m_{D_s^+}+m_{\chi_{c1}(1p)}} - (\varepsilon^* q)\frac{2m_{D_s^+}}{q^2}[A_3(q^2)-A_0(q^2)]q_\mu] \qquad (3)$$

In Eq. (3), $P_\mu = (p + p')_\mu$, $q_\mu = (p - p')_\mu$ and $\varepsilon$ is the polarization vector of the axial vector meson. To have finite results at $q^2 = 0$, the condition of $A_3(0) = A_0(0)$ is required.

The form factor $A_3(q^2)$ can be written as a linear combination of $A_1$ and $A_2$ as follows:

$$A_3(q^2) = \frac{m_{\chi_{c1}(1p)} + m_{D_s^+}}{2m_{D_s^+}} A_1(q^2) - \frac{m_{\chi_{c1}(1p)} - m_{D_s^+}}{2m_{D_s^+}} A_2(q^2) \tag{4}$$

The form factors $V, A_1$ and $A_2$ are calculated by using the following three-point correlation function,

$$\Pi_{\mu\nu} = i^2 \int d^4x\, d^4y\, e^{-ipx} e^{ip'y} \left\langle 0 \left| T\{J_{D_s^+}(y) J_\mu(0) J^\dagger_{\chi_{c1}(1p)}(x)\} \right| 0 \right\rangle \tag{5}$$

where $J_{\chi_{c1}(1p)}(x) = \bar{c}\gamma_\nu\gamma_5 c$ is the interpolating current of the axial meson and $J_{D_s^+}(y) = \bar{c}\gamma_5 s$ is the interpolating current of the pseudoscalar meson. The transition current is, $J_\mu(0) = \bar{s}\gamma_\mu(1 - \gamma_5)c$.

The above correlation function is calculated in two different approaches: first, in the hadron context which is called the phenomenological or physical part and the second, is QCD or theoretical approach, obtained in the quark gluon language. The form factor expressions are determined by equating the corresponding coefficients of the two parts. We use double Borel transformation with respect to p and p′ to suppress the contributions coming from higher states and continuum.

To obtain the physical part, a complete set of intermediate states with the same quantum numbers are inserted in Eq. (5). Therefore we obtain:

$$\Pi_{\mu\nu}(p^2, p'^2, q^2) = \frac{\left\langle 0 \left| J_{D_s^+} \right| D_s^+(p') \right\rangle \left\langle D_s^+(p') \left| J_\mu(0) \right| \chi_{c1}(1p)(p) \right\rangle \left\langle \chi_{c1}(1p)(p) \left| J^\dagger_{\nu,\chi_{c1}(1p)} \right| 0 \right\rangle}{\left(p'^2 - m_{D_s^+}^2\right)\left(p^2 - m_{\chi_{c1}(1p)}^2\right)} +$$
the higher resonances and continuum $\tag{6}$

The matrix elements in Eq. (6) can be parameterized in terms of the leptonic decay constants of $D_s^+$ and $\chi_{c1}(1p)$ mesons as:

$$\left\langle 0 \left| J_{D_s^+} \right| D_s^+(p') \right\rangle = \frac{i f_{D_s^+} m_{D_s^+}^2}{m_c + m_s}, \quad \left\langle 0 \left| J_{\nu,\chi_{c1}(1p)} \right| \chi_{c1}(1p)(p) \right\rangle = f_{\chi_{c1}(1p)} m_{\chi_{c1}(1p)} \varepsilon_\nu \tag{7}$$

To obtain the physical part, Eq. (3) and Eq. (7) are substituted in Eq. (6) and the summation is performed over the polarization of $\chi_{c1}(1p)$ meson.

$$\begin{aligned}
\Pi_{\mu\nu}(p^2, p'^2, q^2) &= \frac{f_{D_s^+} f_{\chi_{c1}(1p)} m_{D_s^+}^2 m_{\chi_{c1}(1p)}}{(m_c + m_s)(p'^2 - m_{D_s^+}^2)(p^2 - m_{\chi_{c1}(1p)}^2)} \Big\{ i\, \epsilon_{\mu\nu\alpha\beta}\, p^\alpha p'^\beta \frac{2V(q^2)}{m_{D_s^+} + m_{\chi_{c1}(1p)}} \\
&\quad - (m_{D_s^+} + m_{\chi_{c1}(1p)})\left(-g_{\mu\nu} + \frac{(P+q)_\mu (P+q)_\nu}{4m_{\chi_{c1}(1p)}^2}\right) A_1(q^2) \\
&\quad + \frac{1}{m_{D_s^+} + m_{\chi_{c1}(1p)}} P_\mu \left(-q_\nu + \frac{pq(P+q)_\nu}{2m_{\chi_{c1}(1p)}^2}\right) A_2(q^2) \\
&\quad + \frac{2m_{D_s^+}}{q^2} q_\mu \left(-q_\nu + \frac{pq(P+q)_\nu}{2m_{\chi_{c1}(1p)}^2}\right) [A_3(q^2) - A_0(q^2)] \Big\}
\end{aligned} \tag{8}$$

To find the expressions for the form factors $V, A_1$ and $A_2$, the coefficients of the Lorentz structures $i\epsilon_{\mu\nu\alpha\beta} p^\alpha p'^\beta$, $g_{\mu\nu}$ and $P_\mu q_\nu$ are required. Therefore, the correlation function is written in terms of the chosen Lorentz structures as follows:

$$\Pi_{\mu\nu}(p^2, p'^2, q^2) = \Pi_V i \epsilon_{\mu\nu\alpha\beta} p^\alpha p'^\beta + \Pi_{A_1} g_{\mu\nu} + \Pi_{A_2} P_\mu q_\nu + \cdots \tag{9}$$

where ... stands for other tensor structures. To find the QCD part of the correlation function, the three–point correlator is determined by using the operator product expansion method (OPE) in the deep Euclidean region $p^2 \ll 4m_c^2$, $p'^2 \ll (m_c^2 + m_s^2)$. Then, the correlation function may appear in perturbative and nonperturbative parts as follows:

$$\Pi_i(p^2, p'^2, q^2) = \Pi_i^{per}(p^2, p'^2, q^2) + \Pi_i^{non-per}(p^2, p'^2, q^2) \tag{10}$$

where we use $i$ to indicate $V$, $A_1$ and $A_2$. The bare loop diagram (Fig. 1 a) is considered for the perturbative part. Only the gluon condensate diagrams are considered as first nonperturbative part (Fig. 2 a, b, c, d, e, f) because the double Borel transformations eliminate the light quark condensates, contributing to the correlation function. Diagrams b, c, d in Fig. 1 shows the light quark condensates.

Double dispersion representation used for the bare–loop contribution is written as:

$$\Pi_i^{per} = -\frac{1}{(2\pi)^2} \int ds' \int ds \frac{\rho_i^{per}(s,s',q^2)}{(s-p^2)(s'-p'^2)} + subtraction\ terms, \tag{11}$$

The quark propagators are replaced by Dirac function, using Cutkosky rules, i.e., $\frac{1}{p^2-m^2} \to -2\pi i \delta(p^2 - m^2)$. Such replacement gives the following inequality:

$$-1 \leq \frac{2ss' + (s+s'-q^2)(-s) + 2s(m_c^2 - m_s^2)}{\lambda^{1/2}(s,s',q^2)\lambda^{1/2}(m_c^2,m_c^2,s)} \leq 1 \tag{12}$$

where $\lambda(a,b,c) = a^2 + b^2 + c^2 - 2ab - 2ac - 2bc$. $s'$ has a lower limit equal to $(m_c + m_s)^2$, and the lower limit of s is also determined from Eq. (12).

Once all calculations are carried out, the spectral densities may be written as following:

$$\rho_V(s, s', q^2) = 4N_c((m_s - m_c)B_2 - 2m_c B_1 - m_c I_0) \tag{13}$$

$$\rho_{A_1}(s, s', q^2) = 2N_c(4(-m_s + m_c)A_1 + 2m_c\Delta' I_0$$
$$+ (-m_s + m_c)I_0 \Delta + 2m_c^2(-2m_c + m_s)I_0$$
$$- m_c(u - 2m_c m_s)I_0) \tag{14}$$

$$\rho_{A_2}(s, s', q^2) = 2N_c((-m_s + m_c)(-A_5 + A_2) - m_c B_2 - m_s B_1) \tag{15}$$

here $u = s + s' - q^2$, $\Delta = s$, $\Delta' = s' + m_c^2 - m_s^2$ and $N_c = 3$ is the number of colors. $B_1, B_2, A_1, A_2, A_5$ and $I_0$ are as follows:

$$I_0(s, s', q^2) = \frac{1}{4\lambda^{1/2}(s,s',q^2)},$$

$$\lambda(s, s', q^2) = s^2 + s'^2 + q^4 - 2sq^2 - 2s'q^2 - 2ss',$$

$$B_1 = \frac{1}{4\lambda^{3/2}}(2s'\Delta - \Delta' u),$$

$$B_2 = \frac{1}{4\lambda^{3/2}}(2s\Delta' - \Delta u),$$

$$A_1 = \frac{1}{8\lambda^{3/2}}\left(\Delta'^2 s + \Delta^2 s' - 4m_c^2 ss' - \Delta\Delta' u + m_c^2 u^2\right),$$

$$A_2 = \frac{1}{4\lambda^{5/2}}(2\Delta'^2 ss' + 6\Delta^2 s'^2 - 8m_c^2 ss'^2 - 6\Delta\Delta' s' u + \Delta'^2 u^2 + 2m_c^2 s' u^2),$$

$$A_5 = \frac{1}{4\lambda^{5/2}}\Big(-6\Delta\Delta' s' u + 6s^2\Delta'^2 - 8s^2 s' m_c^2 + 2u^2 s m_c^2 + u^2\Delta^2 + 2ss'\Delta^2\Big). \tag{16}$$

In order to calculate gluon condensate contributions we should perform proper integrals which are discussed next [20, 24, 26, 30]. We use Fock–Schwinger fixed–point gauge, $x^\mu G_\mu^a = 0$ where $G_\mu^a$ is the gluon field [24, 31-33]. Let us list the necessary integrals to calculate the corresponding diagrams:

$$I_0[a, b, c] = \int \frac{d^4 k}{(2\pi)^4} \frac{1}{[k^2 - m_c^2]^a [(p+k)^2 - m_c^2]^b [(p'+k)^2 - m_s^2]^c},$$

$$I_\mu[a, b, c] = \int \frac{d^4 k}{(2\pi)^4} \frac{k_\mu}{[k^2 - m_c^2]^a [(p+k)^2 - m_c^2]^b [(p'+k)^2 - m_s^2]^c},$$

$$I_{\mu\nu}[a, b, c] = \int \frac{d^4 k}{(2\pi)^4} \frac{k_\mu k_\nu}{[k^2 - m_c^2]^a [(p+k)^2 - m_c^2]^b [(p'+k)^2 - m_s^2]^c}. \tag{17}$$

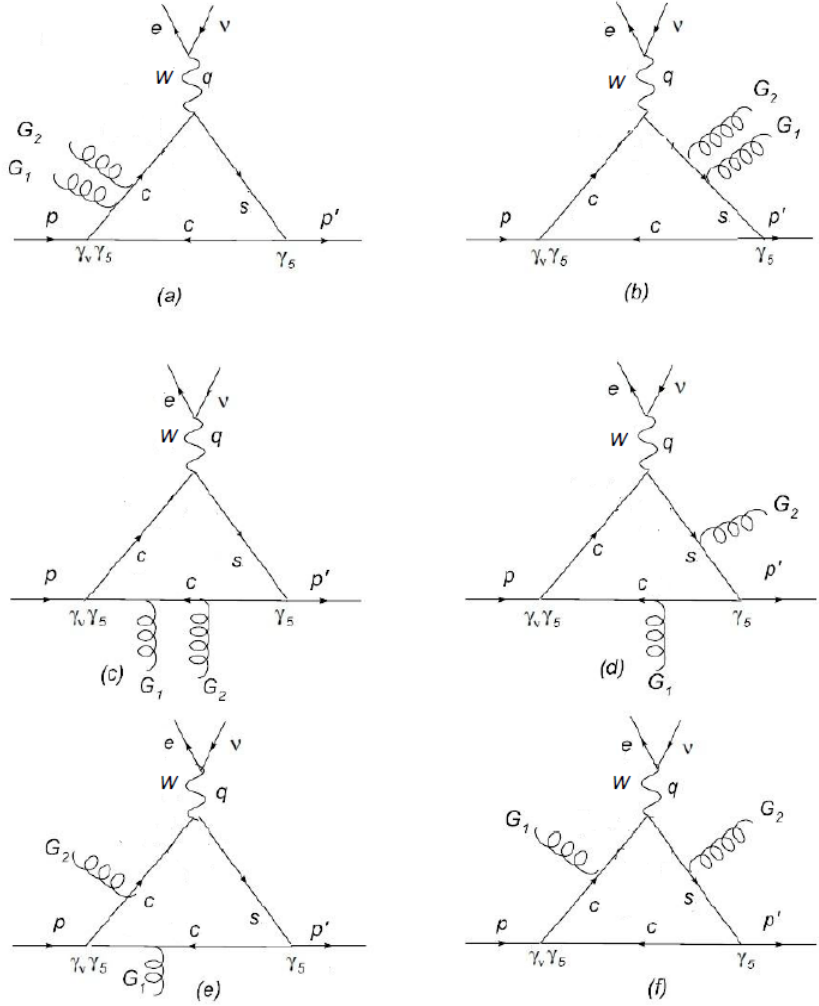

Figure 2: Gluon condensate contributions to $\chi_{c1}(1p) \to D_s^+ e \bar{\nu}$ decay.

These integrals can be solved using Schwinger representation for propagators as follow,

$$\frac{1}{p^2+m^2} = \frac{1}{\Gamma(\alpha)} \int_0^\infty d\alpha\, \alpha^{n-1} e^{-\alpha(p^2+m^2)} \tag{18}$$

The Borel transformation which is proper here is as follows:

$$\hat{B}_{p^2}(M^2) e^{-\alpha p^2} = \delta(1-\alpha M^2) \tag{19}$$

The transformed results of the integrals after solving them and applying double Borel transformations over $p^2$ and $p'^2$, are written as:

$$\begin{aligned}
\hat{I}_0(a,b,c) &= i \frac{(-1)^{a+b+c+1}}{16\pi^2 \Gamma(a)\Gamma(b)\Gamma(c)} (M_1^2)^{2-a-b} (M_2^2)^{2-a-c} U_0(a+b+c-4, 1-c-b), \\
\hat{I}_\mu(a,b,c) &= \tfrac{1}{2}\big[\hat{I}_1(a,b,c) + \hat{I}_2(a,b,c)\big] P_\mu + \tfrac{1}{2}\big[\hat{I}_1(a,b,c) - \hat{I}_2(a,b,c)\big] q_\mu, \\
\hat{I}_{\mu\nu}(a,b,c) &= \hat{I}_6(a,b,c) g_{\mu\nu} + \tfrac{1}{4}(2\hat{I}_4 + \hat{I}_3 + \hat{I}_5) P_\mu P_\nu + \tfrac{1}{4}(-\hat{I}_5 + \hat{I}_3) P_\mu q_\nu \\
&\quad + \tfrac{1}{4}(-\hat{I}_5 + \hat{I}_3) P_\nu q_\mu + \tfrac{1}{4}(-2\hat{I}_4 + \hat{I}_3 + \hat{I}_5) q_\mu q_\nu.
\end{aligned} \tag{20}$$

here:

$$\hat{I}_1(a,b,c) = i\frac{(-1)^{a+b+c+1}}{16\,\pi^2\Gamma(a)\Gamma(b)\Gamma(c)}(M_1^2)^{2-a-b}(M_2^2)^{3-a-c}U_0(a+b+c-5, 1-c-b),$$

$$\hat{I}_2(a,b,c) = i\frac{(-1)^{a+b+c+1}}{16\,\pi^2\Gamma(a)\Gamma(b)\Gamma(c)}(M_1^2)^{3-a-b}(M_2^2)^{2-a-c}U_0(a+b+c-5, 1-c-b),$$

$$\hat{I}_3(a,b,c) = i\frac{(-1)^{a+b+c}}{16\,\pi^2\Gamma(a)\Gamma(b)\Gamma(c)}(M_1^2)^{2-a-b}(M_2^2)^{4-a-c}U_0(a+b+c-6, 1-c-b),$$

$$\hat{I}_4(a,b,c) = i\frac{(-1)^{a+b+c}}{16\,\pi^2\Gamma(a)\Gamma(b)\Gamma(c)}(M_1^2)^{3-a-b}(M_2^2)^{3-a-c}U_0(a+b+c-6, 1-c-b),$$

$$\hat{I}_5(a,b,c) = i\frac{(-1)^{a+b+c}}{16\,\pi^2\Gamma(a)\Gamma(b)\Gamma(c)}(M_1^2)^{4-a-b}(M_2^2)^{2-a-c}U_0(a+b+c-6, 1-c-b),$$

$$\hat{I}_6(a,b,c) = i\frac{(-1)^{a+b+c+1}}{32\,\pi^2\Gamma(a)\Gamma(b)\Gamma(c)}(M_1^2)^{3-a-b}(M_2^2)^{3-a-c}U_0(a+b+c-6, 2-c-b).$$

(21)

In Eqs. (20) and (21), $M_1^2$ and $M_2^2$ are the Borel parameters in the s and s′ channels, respectively, and the function $U_0(\alpha, \beta)$ is defined as follows:

$$U_0(\alpha,\beta) = \int_0^\infty dy(y + M_1^2 + M_2^2)^\alpha y^\beta \exp[-\frac{B_{-1}}{y} - B_0 - B_1 y] \quad (22)$$

$$B_{-1} = \frac{1}{M_1^2 M_2^2}[m_s^2 M_1^4 + m_s^2 M_2^4 + M_1^2 M_2^2(m_c^2 + m_s^2 - q^2)] \quad (23)$$

$$B_0 = \frac{1}{M_1^2 M_2^2}[(m_s^2 + m_c^2)M_1^2 + 2m_c^2 M_2^2] \quad (24)$$

$$B_1 = \frac{m_c^2}{M_1^2 M_2^2} \quad (25)$$

After doing the calculations, the following results are obtained for gluon condensate contributions.

$$\Pi_i^{\langle G^2 \rangle} = i\langle\frac{\alpha_s}{\pi}G^2\rangle\frac{C_i}{24} \quad (26)$$

where expressions for $C_i$ are given in appendix–A.

Double Borel transformations with respect to the $p^2$ ($p^2 \to M_1^2$) and $p'^2$ ($p'^2 \to M_2^2$) are applied on the physical side and QCD side of the correlation function to find the form factors. After matching the coefficient of the Lorentz structures of these two representations of the correlator and doing the continuum subtraction to suppress the higher states and continuum, the following sum rules for the form factors, $A_1$ and $A_2$ are obtained as follows:

$$V = \frac{(m_c + m_s)(m_{\chi_{c1}(1p)} + m_{D_s^+})}{2 f_{D_s^+} f_{\chi_{c1}(1p)} m_{D_s^+}^2 m_{\chi_{c1}(1p)}} e^{\frac{m_{D_s^+}^2}{M_2^2}} e^{\frac{m_{\chi_{c1}(1p)}^2}{M_1^2}}$$

$$\times \left\{ -\frac{1}{4\pi^2} \int_{(m_c+m_s)^2}^{s_0'} ds' \int_{s_L}^{s_0} ds \, \rho_V(s,s',q^2) e^{-\frac{s'}{M_2^2}} e^{-\frac{s}{M_1^2}} + i M_1^2 M_2^2 \langle \frac{\alpha_s}{\pi} G^2 \rangle \frac{C_V}{24} \right\}$$

$$A_1 = \frac{(m_c+m_s)}{f_{D_s^+} f_{\chi_{c1}(1p)} m_{D_s^+}^2 m_{\chi_{c1}(1p)} (m_{\chi_{c1}(1p)} + m_{D_s^+})} e^{\frac{m_{D_s^+}^2}{M_2^2}} e^{\frac{m_{\chi_{c1}(1p)}^2}{M_1^2}}$$

$$\times \left\{ -\frac{1}{4\pi^2} \int_{(m_c+m_s)^2}^{s_0'} ds' \int_{s_L}^{s_0} ds \, \rho_{A_1}(s,s',q^2) e^{-\frac{s'}{M_2^2}} e^{-\frac{s}{M_1^2}} + \right.$$

$$\left. i M_1^2 M_2^2 \langle \frac{\alpha_s}{\pi} G^2 \rangle \frac{C_{A_1}}{24} \right\}$$

$$A_2 = \frac{4 m_{\chi_{c1}(1p)} (m_c+m_s)(m_{\chi_{c1}(1p)} + m_{D_s^+})}{f_{D_s^+} f_{\chi_{c1}(1p)} m_{D_s^+}^2 (3 m_{\chi_{c1}(1p)}^2 + m_{D_s^+}^2 - q^2)} e^{\frac{m_{D_s^+}^2}{M_2^2}} e^{\frac{m_{\chi_{c1}(1p)}^2}{M_1^2}}$$

$$\times \left\{ -\frac{1}{4\pi^2} \int_{(m_c+m_s)^2}^{s_0'} ds' \int_{s_L}^{s_0} ds \, \rho_{A_2}(s,s',q^2) e^{-\frac{s'}{M_2^2}} e^{-\frac{s}{M_1^2}} + \right.$$

$$\left. i M_1^2 M_2^2 \langle \frac{\alpha_s}{\pi} G^2 \rangle \frac{C_{A_2}}{24} \right\} \tag{27}$$

where $s_0$ and $s_0'$ are the continuum thresholds in $\chi_{c1}(1p)$ and $D_s^+$ channels, respectively and $s_L$ is as follows:

$$s_L = \frac{m_c^2 (q^2 - s')^2}{(q^2 - m_c^2)(m_c^2 - s')} \tag{28}$$

Now let us apply the quark-hadron duality assumption to subtract the contributions of the higher states and continuum in Eq. (27):

$$\rho^{higherstates}(s,s') = \rho^{OPE}(s,s') \theta(s - s_0) \theta(s - s_0') \tag{29}$$

The form of Borel transformation which is used is as follows:

$$\hat{B}_{p^2}(M^2) \left\{ \frac{1}{p^2 - m^2} \right\} = -\frac{1}{M^2} e^{-m^2/M^2} \tag{30}$$

The differential decay width $d\Gamma/dq^2$ for the process $\chi_{c1}(1p) \to D_s^+ e \bar{\nu}$ in terms of the form factors is obtained as follows:

$$\frac{d\Gamma}{dq^2 d\cos\theta} = \frac{\sqrt{\lambda}}{256 \pi^3 m_{\chi_{c1}(1p)}^3} |M|^2 \tag{31}$$

$$M = \frac{G_F}{\sqrt{2}} V_{cs} L^\mu H_\mu \tag{32}$$

$$|M|^2 = \frac{G_F^2}{2} |V_{cs}|^2 L^{\mu\nu} H_\mu H_\nu^\dagger \tag{33}$$

$$L^{\mu\nu} H_\mu H_\nu^\dagger = \frac{1}{3}\{\frac{(12m_{\chi_{c1}(1p)}^2 q^2 + \lambda \sin^2\theta)(m_{\chi_{c1}(1p)} + m_{D_s^+})^2}{m_{h_b(1p)}^2} A_1^2 - \frac{2}{m_{\chi_{c1}(1p)}^2}(-m_{D_s^+}^2 + m_{\chi_{c1}(1p)}^2 +$$

$$q^2)\lambda \sin^2\theta\, A_1 A_2 + \frac{1}{m_{\chi_{c1}(1p)}^2 (m_{D_s^+} + m_{\chi_{c1}(1p)})^2} \lambda^2 \sin^2\theta\, A_2^2 + 16\sqrt{\lambda}\, q^2 \cos\theta\, A_1 V\} \quad (34)$$

Where, $\lambda = m_{\chi_{c1}(1p)}^4 + m_{D_s^+}^4 + q^4 - 2m_{\chi_{c1}(1p)}^2 m_{D_s^+}^2 - 2m_{\chi_{c1}(1p)}^2 q^2 - 2m_{D_s^+}^2 q^2$.

Finally, the integration of Eq. (31) over $q^2$ in the interval $0 < q^2 < (m_{\chi_{c1}(1p)} - m_{D_s^+})^2$ is carried out to find the total decay width.

### 3 : Numerical Calculations and results

Considering the expressions for form factors, it is clear that input parameters entering our calculations are gluon condensate, elements of the CKM matrix $V_{cs}$, leptonic decay constants, $f_{\chi_{c(1p)}}$ and $f_{D_s^+}$, quark and meson masses, continuum thresholds $s_0$ and $s_0'$, as well as the Borel parameters $M_1^2$ and $M_2^2$. The values of these parameters are chosen to be:

$\langle \frac{\alpha_s}{\pi} G^2 \rangle = 0.012\, GeV^4$ [7], $|V_{cs}| = 0.957 \pm 0.11$ [34], $f_{D_s^+} = 0.274 \pm 0.013\, GeV$ [35], $f_{\chi_{c(1p)}} = 344 \bullet \pm 27\, MeV$ [36], $m_c = 1.275 \pm 0.015\, GeV$, $m_s(1\, GeV) \simeq 142\, MeV$ [37], $m_{D_s^+} = 1.968\, GeV$ [34], $m_{\chi_{c(1p)}} = (3.51066 \pm 0.00007)\, GeV$ [36]. Form factors contain four parameters: Borel mass squares $M_1^2$ and $M_2^2$ and continuum thresholds, $s_0$ and $s_0'$. These are mathematical parameters, so the physical quantities, such as form factors, should not depend upon them. The parameters $s_0$ and $s_0'$, the continuum thresholds of $\chi_c(1p)$ and $D_s^+$ mesons, are determined from the conditions that guarantee the sum rules to have the best stability in the allowed $M_1^2$ and $M_2^2$ region. The value of the continuum thresholds, $s_0'$ calculated from the two–point QCD sum rules is taken to be $s_0' = 6 - 8\, GeV^2$ [38] and $s_0 = 16 \pm 2\, GeV^2$ [39]. The allowed regions of $M_1^2$ and $M_2^2$ for the form factors are determined from the condition that guarantees the best stability for the form factors. This condition is satisfied in the regions $10\, GeV^2 \leq M_1^2 \leq 15\, GeV^2$ and $20\, GeV^2 \leq M_2^2 \leq 30\, GeV^2$ as shown in figure 3. The values of the form factors at $q^2 = 0$ are shown in Table 1.

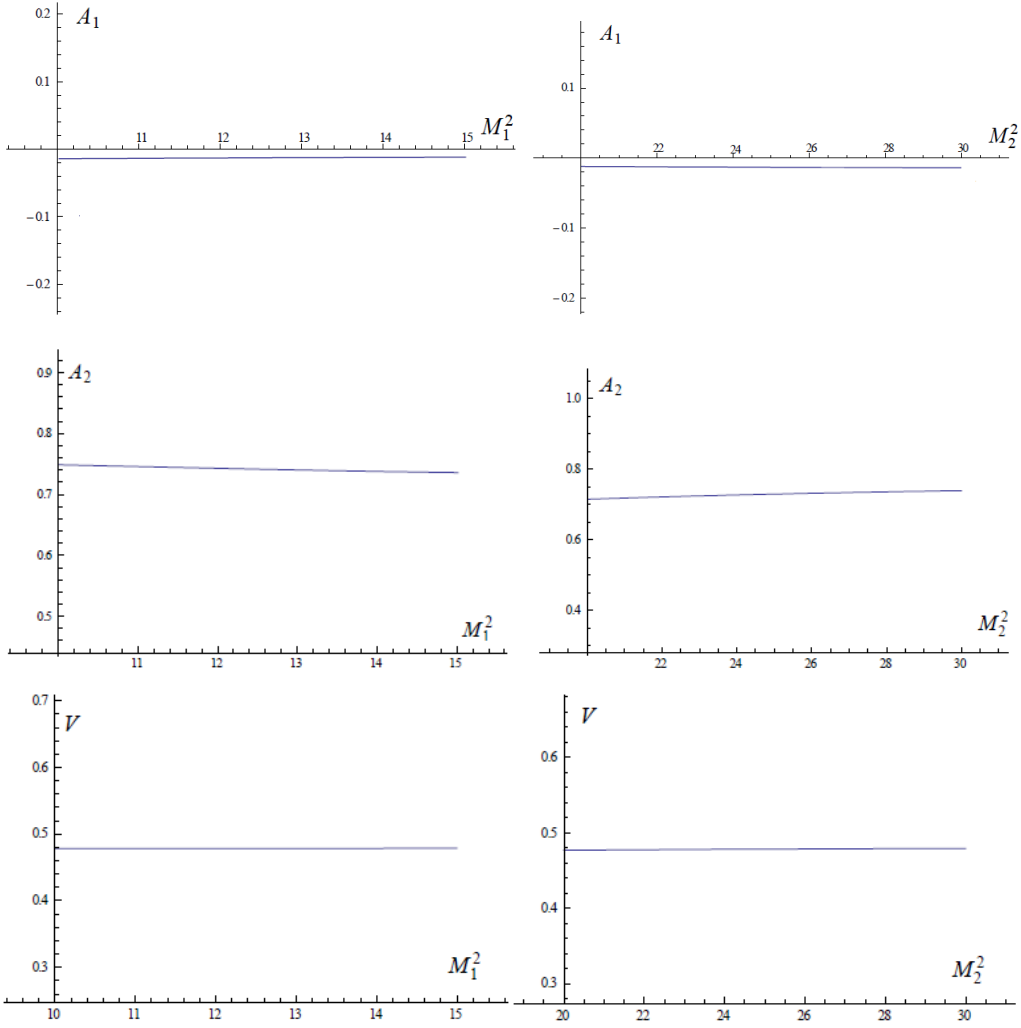

Fig 3: Dependence of the form factors on Borel parameters.

|  | $\chi_{c1}(1p) \to D_s^+ e\bar{\nu}$ |
|---|---|
| $V(0)$ | 0.479 |
| $A_1(0)$ | $-0.017$ |
| $A_2(0)$ | 0.733 |

Table 1: The values of the form factors at $q^2 = 0$, for $M_1^2 = 12.5\ GeV^2$, $M_2^2 = 25\ GeV^2$.

Since the sum rules for the form factors are truncated at some points, in order to extend our calculations to the full physical range, i.e., the region $0 \leq q^2 \leq 2.38\ GeV^2$, we use suitable parameterization for the form factors. Our numerical calculations shows that the best parameterization of the form factors with respect to $q^2$ are as follows:

$$f_i(q^2) = \frac{a}{(1-\frac{q^2}{m_{fit}^2})} + \frac{b}{(1-\frac{q^2}{m_{fit}^2})^2} \qquad (35)$$

The values of the parameters, $a$ and $b$ are given in Table 2:

|       | $a$    | $b$   | $m_{fit}$ (GeV) |
|-------|--------|-------|-----------------|
| $V$   | 0.25   | 0.22  | 2.1             |
| $A_1$ | 0.50   | -0.51 | 2.6             |
| $A_2$ | -51.20 | 51.88 | 2.4             |

Table 2: Parameters in the fit function of the form factors, for $M_1^2 = 12.5 \ GeV^2$, $M_2^2 = 25 \ GeV^2$.

Performing the integration over $q^2$ in Eq. (31) in the interval $0 < q^2 < (m_{\chi_{c(1p)}} - m_{D_s^+})^2$, we get the expression for the total decay width. Our calculated value of the branching fraction is presented in Table 3.

| Decay | Branching fraction |
|-------|--------------------|
| $\chi_{c1}(1p) \to D_s^+ e \bar{\nu}$ | $8.67 \times 10^{-10}$ |

Table 3: Branching fraction of $\chi_{c1}(1p) \to D_s^+ e \bar{\nu}$.

## 4 : Conclusion

In the present work, we studied the rare $\chi_{c1}(1p) \to D_s^+ e \bar{\nu}$ decay in the context of the three-point QCD sum rules. Taking into account the two-gluon corrections to the correlation function as a first nonperturbative contribution, we obtained the form factors. Implementing our findings, we used proper parametrization for the form factors to calculate the branching fraction. The present predictions can be confirmed by the experimental data in the future.

# Appendix-A

$$C_V = 16m_c\hat{I}_0(1,2,2) - 96m_c^3\hat{I}_0(1,4,1) + 64\,m_c^3\hat{I}_0(2,3,1) + 16m_c\hat{I}_0(3,1,1) - 16\,m_c^3\hat{I}_0(3,1,2)$$
$$- 16m_c\hat{I}_0^{[0,1]}(3,1,2) - 16m_c\hat{I}_0^{[1,0]}(3,2,1) - 32m_c^3\hat{I}_0^{[0,1]}(3,2,2) - 16m_c^3\hat{I}_0^{[1,0]}(3,2,2)$$
$$+ 16m_c\hat{I}_0^{[1,1]}(3,2,2) + 32m_c\hat{I}_1(1,2,2) - 96m_c\hat{I}_1(1,3,1) - 192m_c^3\hat{I}_1(1,4,1) - 32m_c\hat{I}_1(2,2,1)$$
$$+ 32m_c\hat{I}_1^{[1,0]}(2,3,1) + 16m_c\hat{I}_1(3,1,1) - 48m_c^3\hat{I}_1(3,1,2) - 16m_c\hat{I}_1^{[0,1]}(3,1,2)$$
$$- 32m_c\hat{I}_1^{[1,0]}(3,2,1) - 64m_c^3\hat{I}_1^{[0,1]}(3,2,2) - 32m_c^3\hat{I}_1^{[1,0]}(3,2,2) + 32m_c\hat{I}_1^{[1,1]}(3,2,2)$$
$$+ 16m_c\hat{I}_2(1,2,2) - 96m_c^3\hat{I}_2(1,4,1) + 64m_c^3\hat{I}_2^{[0,1]}(2,3,1) - 16m_c^3\hat{I}_2(3,1,2)$$
$$- 16m_c\hat{I}_2^{[0,1]}(3,1,2) - 64m_c^3\hat{I}_2(3,2,1) - 32m_c^3\hat{I}_2^{[0,1]}(3,2,2) - 16m_c^3\hat{I}_2^{[1,0]}(3,2,2)$$
$$+ 16m_c\hat{I}_2^{[1,1]}(3,2,2)$$

$$C_{A_1} = -8m_c\hat{I}_0(1,1,2) - 16m_c\hat{I}_0(1,2,1) - 16m_c^3\hat{I}_0(1,2,2) + 8m_c\hat{I}_0^{[0,1]}(1,2,2) + 96m_c^3\hat{I}_0(1,3,1)$$
$$- 48m_c\hat{I}_0^{[0,1]}(1,3,1) + 96m_c^5\hat{I}_0(1,4,1) - 48m_c^3\hat{I}_0^{[0,1]}(1,4,1) + 16m_c^3\hat{I}_0(2,2,1)$$
$$+ 16m_c\hat{I}_0^{[0,1]}(2,2,1) + 32m_c^3\hat{I}_0^{[0,1]}(2,3,1) + 16m_c^3\hat{I}_0^{[1,0]}(2,3,1) - 16m_c\hat{I}_0^{[1,1]}(2,3,1)$$
$$+ 8m_c\hat{I}_0^{[0,1]}(3,1,1) - 8m_c\hat{I}_0^{[1,0]}(3,1,1) + 24m_c^5\hat{I}_0(3,1,2) + 8m_c^3\hat{I}_0^{[0,1]}(3,1,2)$$
$$+ 8m_c^3\hat{I}_0^{[1,0]}(3,1,2) - 8m_c\hat{I}_0^{[1,1]}(3,1,2) + 32m_c^3\hat{I}_0^{[0,1]}(3,2,1) - 24m_c\hat{I}_0^{[1,1]}(3,2,1)$$
$$+ 8m_c\hat{I}_0^{[2,0]}(3,2,1) + 32m_c^5\hat{I}_0^{[0,1]}(3,2,2) - 16m_c^3\hat{I}_0^{[0,2]}(3,2,2) + 16m_c^5\hat{I}_0^{[1,0]}(3,2,2)$$
$$- 24m_c^3\hat{I}_0^{[1,1]}(3,2,2) + 8m_c\hat{I}_0^{[1,2]}(3,2,2) + 32m_c\hat{I}_6(1,2,2) - 192m_c^3\hat{I}_6(1,4,1)$$
$$- 64m_c\hat{I}_6^{[0,1]}(2,3,1) - 64m_c\hat{I}_6^{[1,0]}(2,3,1) - 64m_c^3\hat{I}_6^{[0,1]}(3,2,2) - 32m_c^3\hat{I}_6^{[1,0]}(3,2,2)$$
$$+ 32m_c\hat{I}_6^{[1,1]}(3,2,2)$$

$$C_{A_2} = -16m_c^3\hat{I}_0(3,1,2) + 16m_c^3\hat{I}_0(3,2,1) - 4m_c\hat{I}_1(1,2,2) + 24m_c\hat{I}_1(1,3,1) + 24m_c^3\hat{I}_1(1,4,1) - 8m_c\hat{I}_1(2,2,1)$$
$$- 16m_c^3\hat{I}_1(2,3,1) + 8m_c\hat{I}_1^{[1,0]}(2,3,1) - 4m_c\hat{I}_1(3,1,1) - 8m_c\hat{I}_1(3,1,2) + 16m_c^3\hat{I}_1(3,2,1)$$
$$+ 8m_c\hat{I}_1^{[1,0]}(3,2,1) + 8m_c^3\hat{I}_1^{[0,1]}(3,2,2) + 4m_c^3\hat{I}_1^{[1,0]}(3,2,2) - 4m_c\hat{I}_1^{[1,1]}(3,2,2) - 4m_c\hat{I}_2(1,2,2)$$
$$+ 24m_c\hat{I}_2(1,3,1) + 24m_c^3\hat{I}_2(1,4,1) - 8m_c\hat{I}_2(2,2,1) - 16m_c^3\hat{I}_2(2,3,1) + 8m_c\hat{I}_2^{[1,0]}(2,3,1)$$
$$- 4m_c\hat{I}_2(3,1,1) - 8m_c^3\hat{I}_2(3,1,2) + 16m_c^3\hat{I}_2(3,2,1) + 8m_c\hat{I}_2^{[1,0]}(3,2,1) + 8m_c^3\hat{I}_2^{[0,1]}(3,2,2)$$
$$+ 4m_c^3\hat{I}_2^{[1,0]}(3,2,2) - 4m_c\hat{I}_2^{[1,1]}(3,2,2) + 8m_c\hat{I}_3(1,2,2) - 48m_c^3\hat{I}_3(1,4,1) - 16m_c\hat{I}_3^{[0,1]}(2,3,1)$$
$$- 16m_c\hat{I}_3^{[1,0]}(2,3,1) - 16m_c^3\hat{I}_3^{[0,1]}(3,2,2) - 8m_c^3\hat{I}_3^{[1,0]}(3,2,2) + 8m_c\hat{I}_3^{[1,1]}(3,2,2)$$
$$- 8m_c\hat{I}_5(1,2,2) + 48m_c^3\hat{I}_5(1,4,1) + 16m_c^3\hat{I}_5^{[0,1]}(2,3,1) + 16m_c^3\hat{I}_5^{[1,0]}(2,3,1)$$
$$+ 16m_c^3\hat{I}_5^{[0,1]}(3,2,2) + 8m_c^3\hat{I}_5^{[1,0]}(3,2,2) - 8\,m_c\hat{I}_5^{[1,1]}(3,2,2)$$

Where:

$$\hat{I}_n^{[i,j]}(a,b,c) = (M_1^2)^i (M_2^2)^j \frac{d^i}{d(M_1^2)^i} \frac{d^j}{d(M_2^2)^j} [(M_1^2)^i (M_2^2)^j \hat{I}_n(a,b,c)]$$